\documentclass[aps,prl,twocolumn,english,showpacs,letterpaper,superscriptaddress]{revtex4-1}

\usepackage[dvips]{epsfig}
\usepackage[colorlinks=true,citecolor=blue,linkcolor=blue]{hyperref}

\usepackage{amssymb}
\usepackage{graphicx}
\usepackage{subfigure}
\usepackage{array}

\usepackage{epsfig}

\usepackage{amsmath}
\usepackage{color}
\usepackage{float}
\usepackage{mathrsfs}
\usepackage{indentfirst}
\usepackage{textcomp}
\usepackage{comment}
\usepackage{mathtools}

\newcommand{\ignore}[1]{}
\newcommand{\mater}{ZrTe$_5$}

\newcommand{\COMMENTED}[1]{}

\newcommand{\titledef}{Topological phase transition and phonon-space Dirac topology surfaces in ZrTe$_5$}

\begin{document}

\title{\titledef}
\author{Niraj Aryal}
\email{naryal@bnl.gov}
\affiliation{Condensed Matter Physics and Materials Science Division, Brookhaven National Laboratory, Upton, New York 11973, USA}
\author{Xilian Jin}
\affiliation{Condensed Matter Physics and Materials Science Division, Brookhaven National Laboratory, Upton, New York 11973, USA}
\affiliation{College of Physics, Jilin University, Changchun, Jilin 130012, P.R. China}
%\affiliation{College of Physics, State Key Laboratory of Superhard Materials, Jilin University, Changchun 130012, China}
\author{Q. Li}
\affiliation{Condensed Matter Physics and Materials Science Division, Brookhaven National Laboratory, Upton, New York 11973, USA}
\author{A. M. Tsvelik}
\affiliation{Condensed Matter Physics and Materials Science Division, Brookhaven National Laboratory, Upton, New York 11973, USA}
\author{Weiguo Yin}
\email{wyin@bnl.gov}
\affiliation{Condensed Matter Physics and Materials Science Division, Brookhaven National Laboratory, Upton, New York 11973, USA}

\date{\today}

\begin{abstract}
We use first-principles methods to reveal that in {\mater}, a layered van der Waals material like graphite, atomic displacements corresponding to five of the six zone-center A$_g$ (symmetry-preserving) phonon modes can drive a topological phase transition from strong to weak topological insulator with a Dirac semimetal state emerging at the transition, giving rise to a Dirac topology surface in the multi-dimensional space formed by the A$_g$ phonon modes. This implies that the topological phase transition in {\mater} can be realized with many different settings of external stimuli that are capable of penetrating through the phonon-space Dirac surface without breaking the crystallographic symmetry. Furthermore, we predict that domains with effective mass of opposite signs can be created by laser pumping and will host Weyl modes of opposite chirality propagating along the domain boundaries. Studying phonon-space topology surfaces provides a new route to understanding and utilizing the exotic physical properties of ZrTe$_5$ and related quantum materials.
\end{abstract}

%\pacs{}

\maketitle

%\section{Introduction}
\emph{Introduction---}The prediction and subsequent verification of the first topological insulators about a decade ago~\cite{QSHGrapheneKaneMele2005,ExperimentQSHKonig2007,QSH_HeTeBernevigHughesZhang2006,3DTIFuKaneMele2007,zhang2009topological,xia2009observation}
%cite Kane,Mele's few papers and Bernevig/Zhang's paper
marks a watershed moment in modern condensed matter physics and materials science.
Quite a few forms of topological materials such as topological insulators~\cite{TISmB6Kim2014,zhang2009topological,RevModPhysBansil2016}, Dirac semimetals~\cite{3DDiracA3BiWang2012,3DDiracCd3As2Wang2013,3DDiracCd3As2ExptNeupane2014}, and Weyl semimetals
\cite{WeylPyrochloreVishwanath2011,FermiarcsTaAs_HuangNature2015,WeylTaAsExptDing2015} together with many novel physical properties have been discovered ever since. Some examples are the existence of the symmetry protected surface states first predicted in ~\cite{Volkov1985,Kusmartsev1985}, exotic transport properties in the presence of the electric and magnetic fields, suppressed back-scattering, and a materialized playground to test fundamental theories governing the early universe~\cite{Hasan_Kane_RMP,FermiarcsTaAs_XuScience2015,ABJAnomalyInCrystalsNielsonNinomiya1983,NielsenNinomiya1981,NerstLatticeModel_PRB_2016,Armitage_Vishwanath_RMP2018}.
A key question in this emerging field is how to drive phase transitions between different topological states, as requested for energy and quantum information applications {\it etc.}~\cite{UltrafastSymmetrySwitchWeyl_SieNature2019,SymmetrySwitchMoTe2_ZhangPRX2019,Crossover3Dto2DTI_ZhangNature2010,StrainInducedTopology_Bi2Se3_PRB2011,TopoPhaseTran_Bi2Se3_Doping_Vanderbilt_PRB2013,TopoTransBiSb_Nakamura_PRB2011}.

Zirconium pentatelluride {\mater}, a layered van der Waals material like graphite, has recently been found to hold a unique position among materials with topological phase transition (TPT). This material, which has baffled physicists for decades by its anomalous transport properties ~\cite{ThermopowerZrTe5_Jones_SSC_1982,GiantResistivityAnomalyZrTe5_Okada_JPSJ_1980,EnhancementResistiveAnomalyZrTe5_Tritt_PRB_1999},
%Historically, ZrTe5 was attractive becacue tere was enhancement of reisitivity at around 150 K which is an indication of phase transition or the change of the type of the carrier(??this is what 1982 paper says).
%However, no phase transition has been observed yet.
has once again attracted intense research interest due to its novel and ambiguous topological behavior.
The monolayer of this material is predicted to be in a quantum spin Hall (QSH) phase.
Whereas, the bulk sample is found to be in close proximity to the phase boundary
between strong topological insulator (STI) and weak topological insulator (WTI)~\cite{QSHZrTe5_PRX_Weng_Dai_2014}
and hosts a chiral magnetic effect on electron transport~\cite{CME_ZrTe5_QiangLiNature2016}.
Thus, slight changes in the lattice parameters, which could happen due to different sample growth conditions or other external perturbations such as strain and temperature, allocate the system to either STI or WTI phase~\cite{ZrTeARPES_Manzoni_PRL2016,ZrTeARPES_WeakTI_XiongPRB2017,StrainTunedTopology_Mutch_Science2019,TempDrivenTopology_Xu_PRL2018}.
%A lot of studies have been performed recently regarding the topological phases of van-der Waals layered material ZrTe$_5$.
%For example, some angle-resolved photoemission spectroscopy (ARPES) studies claim to have observed the STI phase of this material~\cite{ZrTeARPES_Manzoni_PRL2016}, whereas others claim to have observed the WTI phase~\cite{ZrTeARPES_Moreschini_PRB2016}, although different measurements agree on the general electronic dispersion of this material, they contradict on the nature of the topological phase which arises from the very-low energy behaviour.
Very recently, ultrafast photoinduced TPT in {\mater} has been reported~\cite{Vaswani2019light,PrivateComm}. Unlike the ultrafast photoinduced TPT in another layered van der Waals material (W,Mo)Te$_2$~\cite{UltrafastSymmetrySwitchWeyl_SieNature2019,SymmetrySwitchMoTe2_ZhangPRX2019},
which was driven by a change in the lattice symmetry from non-centrosymmetric T$_d$ to centrosymmetric 1T$^\prime$ via light pulses induced interlayer shear strain, the one in {\mater} preserves the lattice symmetry: one Raman-active $A_g$ optical phonon mode was excited by intense $\sim 1.2$THz laser pulses in the optical measurements~\cite{Vaswani2019light}, on the other hand, while the MeV ultrafast electron diffraction (UED)~\cite{PrivateComm} used 800nm laser pulses, which do not correspond to any $A_g$ phonon mode and likely excite a combination of $A_g$ phonon modes. With avoiding the complexities and relaxation phenomena associated with the crystallographic phase transition, such TPT could have certain advantages for technological applications. It calls for a timely understanding of its mechanism and in particular the questions as to how many symmetry-preserving phonon modes can drive the TPT individually or jointly. Importantly, with more modes not breaking the crystallographic symmetry, the metastable state created by the laser pumping will likely consist of domains with different phases due to those different modes of lattice distortions and novel effects are expected to take place on the domain boundaries.

In this paper we address these questions by systematically studying the lattice-symmetry-preserving TPT in {\mater}. We use both first-principles calculations and a derived effective Hamiltonian to analyze the electron and phonon band structures. We find that the atomic displacement patterns corresponding to five out of the six A$_g$ modes can drive the TPT from STI to WTI. At the transition point for each mode, the system becomes a Dirac semimetal (DSM), giving rise to a Dirac topology surface in the 6-dimensional (6D) space formed by the A$_g$ Raman phonon modes. Our results indicate that TPT in ZrTe$_5$ can be realized with many different settings of external stimuli that are capable of penetrating through the Dirac surface. An immediate application is using selective terahertz optical pumps to induce resonant response of one of the A$_g$ modes, in which the incident photon fluence is the parameter for controlling the penetration through the Dirac surface. Furthermore, we predict that the domains with the mass terms of opposite sign will have the Weyl modes of opposite chirality propagating along the domain boundaries. In a broader sense, the concept of phonon-space Dirac topology surface can be readily generalized to include other symmetry-breaking Raman-active or infrared-active phonon modes or other kinds of topology such as Weyl semimetal state. Studying phonon-space topology surfaces provides a new generic route to understanding and utilizing the exotic physical properties of ZrTe$_5$ and related quantum materials.

\ignore{The organization of this manuscript is as follows.
In Sec.~\ref{Sec:Computation}, we present the details of our computational methods.
Then in Sec.~\ref{Sec:Results}, we present our results and and discuss them in details.
%In Sec.~\ref{Sec:Discussion}, we present our discuss our results and their implications to the past and future experiments.
Finally, in Sec.~\ref{Sec:Conclusion}, we present our conclusion
and future outlook.
%\textcolor{red}{talk about IR modes and chiral magnetic effect, cite Qiang Li's paper}
}

%\section{Results and Discussion}
%\label{Sec:Results}

%\subsection{Structural relaxation and electronic bands}

\begin{figure}[b]
    \begin{center}
        \includegraphics[width=\columnwidth]{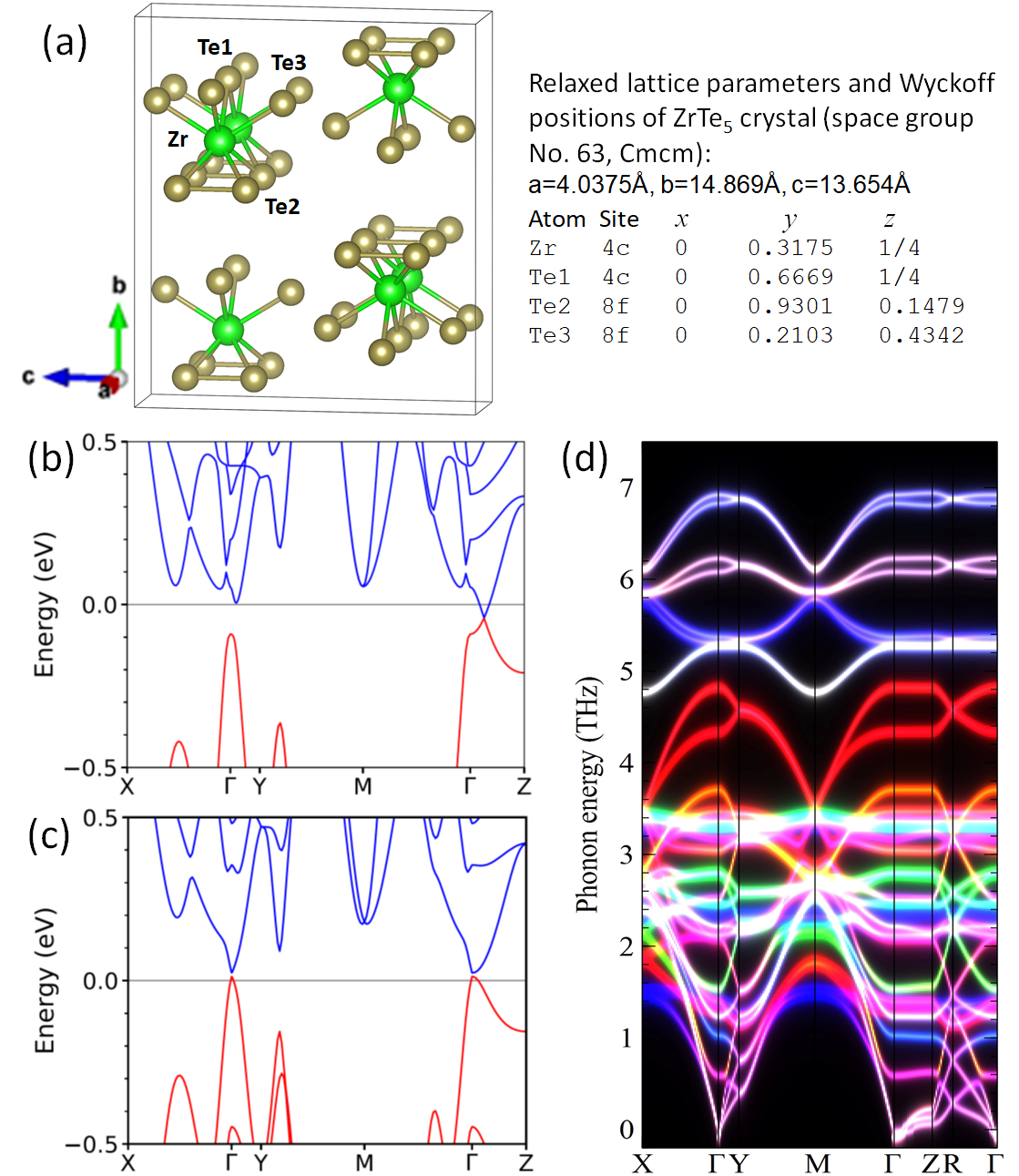}
    \end{center}
    \caption{
           (a) The crystal structure of {\mater} and the relaxed structural data. Calculated band structure (b) without and (c) with the inclusion of SOC, where the labels of the high symmetry points follow Ref.~\onlinecite{QSHZrTe5_PRX_Weng_Dai_2014}. (d) Atom projected phonon dispersion of ZrTe$_5$, where the colors denote contributions from the Zr (white), Te1 (green), Te2 (blue), and Te3 (red) atoms.}
        \label{fig:Relaxed_Expt_Bands}
\end{figure}

\emph{Crystal, electronic, and phonon structures---}The crystal structure of {\mater} and the relaxed structural data are presented in Fig.~1(a) (see Supplemental Note 1~\cite{Supplementary}). The electronic band structures without and with spin-orbit coupling (SOC) being included are shown in Fig.~1(b) and 1(c), respectively,
%The band structure obtained by using the experimental lattice parameters is shown in the Supplementary Material
which overall agree with previous calculations~\cite{QSHZrTe5_PRX_Weng_Dai_2014}.
In the absence of SOC, we see a crossing between the valence and conduction bands along the
$\Gamma$-$Z$ line, which is however gapped by the SOC. The small gap size of 12 meV is similar to the value reported in the literature.
%Since the system is very close to the topological phase boundary, it is important that the relaxed structure gives the correct topological phase of the system as it could affect the conclusions.
From Wannier function tight binding analysis, we find that the SOC affects mainly the on-site Hamiltonian matrix elements (i.e. between different orbitals of the same atom) in terms of $\lambda \mathbf{L} \cdot \mathbf{S}$, where $\lambda$ was found to be 0.007 eV for Zr atoms and 0.36 eV for Te atoms. We find that both the relaxed and and the experimentally observed structures are in the STI phase.

%\subsection{Phonon band structure}
The atom-projected phonon band structure of ZrTe$_5$ is shown in Fig.~1(d). Overall, the Zr-derived $\Gamma$-point vibration modes (above 5HTz in white color) are harder than the Te-derived. There are 36 phonon bands corresponding to 12 atoms (two formula units) in primitive unit cell, including 13 infrared-active and 18 Raman-active optical modes (See Supplemental Note~3~\cite{Supplementary}). The Raman- and infrared-active modes preserve and break the inversion symmetry, respectively. It is thus interesting to observe photoinduced TPT by preserving and breaking inversion symmetry in this material; they may drive the system to Dirac and Weyl semimetal phases, respectively. As an essential first step, we focus on the full crystalline symmetry protecting A$_g$ Raman phonon modes.

There are six A$_g$ modes in total, since the Cmcm space-group symmetry of ZrTe$_5$ crystal contains six independent variables in the atomic positions [Fig.~\ref{fig:Relaxed_Expt_Bands}(a)].
%, whose variation does not change the crystal symmetry and will not introduce additional Bragg peaks in diffraction experiments.
Specifically, the Zr and Te1 atoms move only along the $\textbf{b}$-direction whereas Te2 and Te3 atoms move only in the $\textbf{b-c}$ plane.
These modes are referred to as A$_g$-6~\cite{Vaswani2019light}, A$_g$-22, A$_g$-25, A$_g$-27, A$_g$-29 and A$_g$-36 based on the energy ranking shown in Supplemental Table~S2. Their atomic displacement vectors are shown in Supplemental Fig.~S2. $Q$, the normal coordinate of the phonon modes, is defined in Supplemental Note~2~\cite{Supplementary}. The energy cost as a function of $Q$ reveals that the system is in the harmonic regime for all the mode displacements considered (Supplemental Fig.~S3).

\begin{figure}[t]
    \includegraphics[width=\columnwidth]{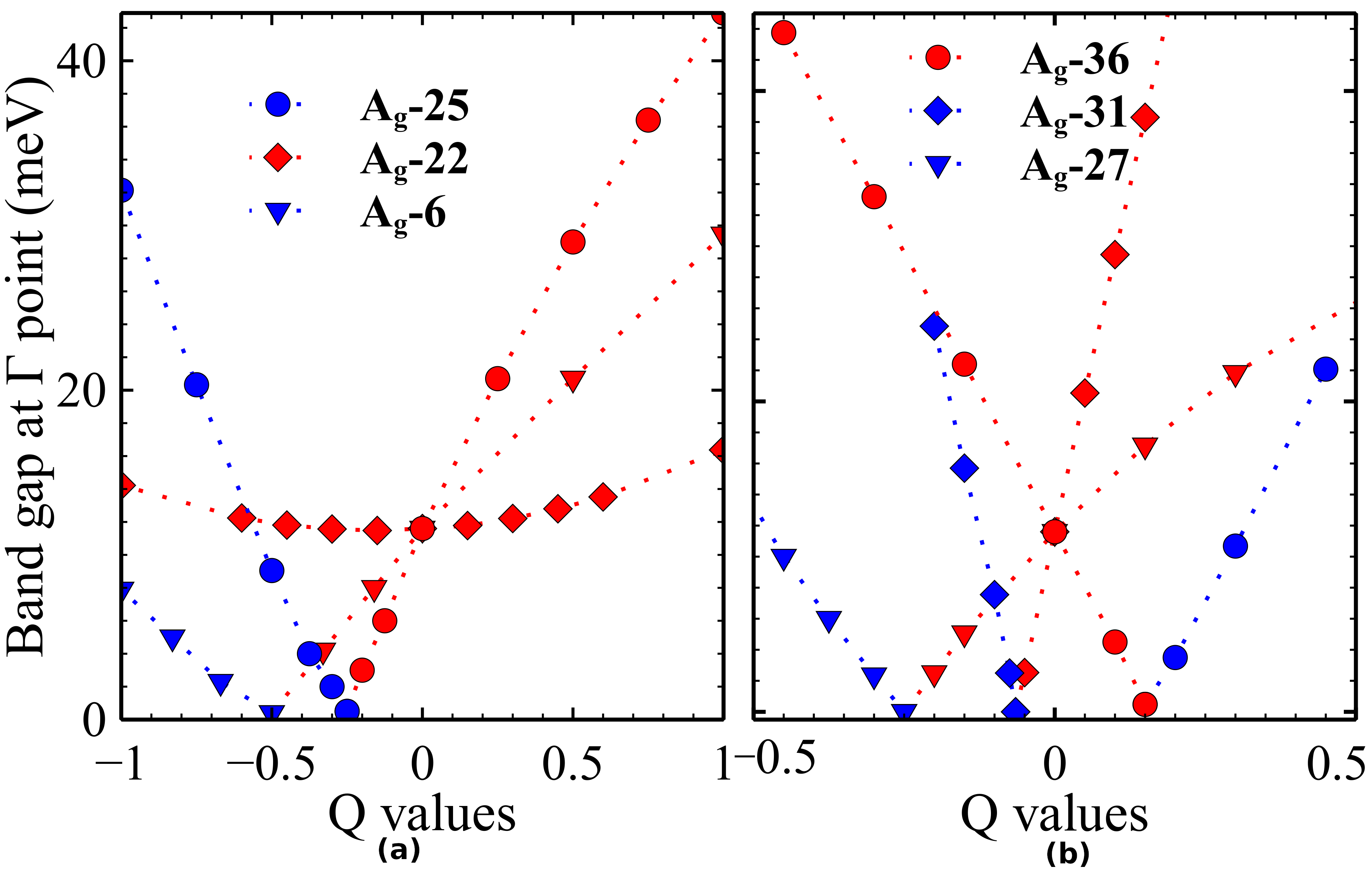}
    \caption{Evolution of the band gap as a function of the lattice displacement factor for (a) lower three A$_g$ modes i.e. A$_g$-6, A$_g$-22 and A$_g$-25 modes and (b) remaining three A$_g$ modes i.e.  A$_g$-27, A$_g$-31 and A$_g$-36 modes.
    The red (blue) dots denote that the system is in STI (WTI) phase for the corresponding value of the lattice displacement factor.
    }
        \label{fig:Bandgap_displacement_summary}
\end{figure}

%\subsection{Topological phase transition}
\emph{Topological phase transition---}The $Z_2$ invariant for each A$_g$ mode at different $Q$ values of the normal coordinates is used to infer about the band topological property. We found that except the A$_g$-22 mode, the other five A$_g$ Raman modes can drive a STI-WTI phase transition. As summarized in
Fig.~\ref{fig:Bandgap_displacement_summary}, the transition is characterized by the closing of the $\Gamma$-point band gap.

\begin{figure}[thb]
    \begin{center}
       \includegraphics[width=\columnwidth]{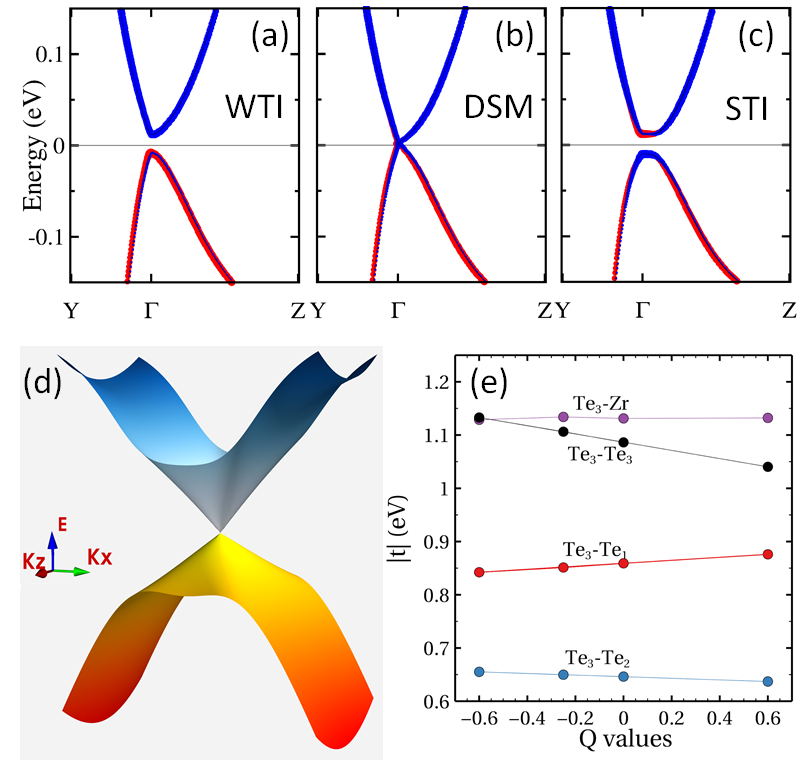}
    \end{center}
    \caption{Evolution of the band structure and orbital content of the bands forming the Dirac cone around the $\Gamma$ point for different values of the normal coordinate $Q$ corresponding to the A$_g$-27 Raman-active phonon mode. The $Q$ values are (a) $-0.6$, (b) $-0.25$, and (c) 0.3. The red (blue) dots show the proportion of the Te4 (Te3) $5p$ orbitals. (d) The 3D view of the Dirac cone on the $k_x-k_z$ plane for $Q=-0.25$. (e) The variation of the hopping strengths between the nearest neighbour atoms as a function of $Q$ for A$_g$-27 mode.
    }
        \label{fig:Ag_27bands}
\end{figure}

%\emph{Mode 27---}
For example, the A$_g$-27 mode [see Supplemental Fig.~S2(d)]---the outstanding red band with frequency of 4.33 THz at $\Gamma$ point in Fig.~\ref{fig:Relaxed_Expt_Bands}(d)---is dominated by the displacement of the Te3 atoms in the $\textbf{b-c}$ plane. The system goes from STI to WTI for $Q<-0.25$, which corresponds to $\sim 0.01$~{\AA} displacement of Te3 atoms in the $\textbf{b-c}$ plane.
%This mode was shown to be the main component of the 800nm laser pulses induced lattice displacement that led to TPT in {\mater} during the UED experiment~\cite{PrivateComm}.
In Fig.~\ref{fig:Ag_27bands}, we show the evolution of the bands for different values of $Q$ corresponding to the A$_g$-27 mode. In the WTI phase [Fig.~\ref{fig:Ag_27bands}(a)], the valence and conduction bands are mainly composed of Te3 and Te2 $5p$ orbitals. In the STI phase [Fig.~\ref{fig:Ag_27bands}(c)], there exists band inversion between these orbitals in the vicinity of the $\Gamma$ point. In between, the gap closes [Fig.~\ref{fig:Ag_27bands}(b)] and a Dirac cone forms at $\Gamma$ point [Fig.~\ref{fig:Ag_27bands}(d)].
Similar discussions for the other A$_g$ phonon modes are shown in Supplemental Note~3~\cite{Supplementary}.

To get more insight into the TPT, we investigate how the nearest (nn) and the next nearest (nnn) neighbour hopping strengths vary as a function of $Q$ for the A$_g$-27 mode. As shown in Fig.~\ref{fig:Ag_27bands}(e), we find that the hoppings between the Te3-Te3 and Te3-Te1 nn are the ones most affected. Other hopping terms are not much affected. Surprisingly, the Te3-Te2 and Te3-Zr nn hoppings are constant even though their relative distance change as a function of $Q$. This implies that the TPT can be induced by varying the Te3-Te3 nn hopping.


\begin{figure*}[t]
    \begin{center}
       \includegraphics[width=0.9\textwidth]{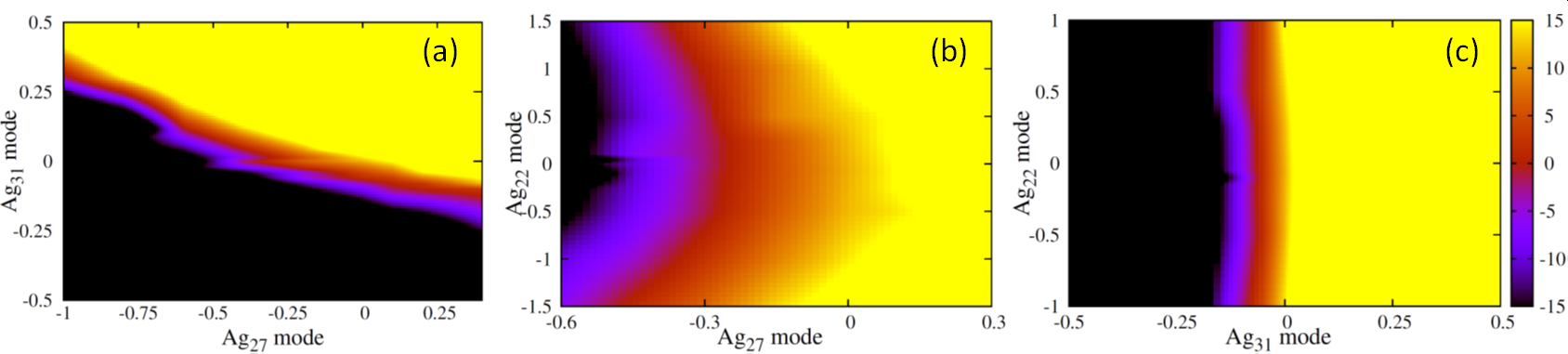}
\ignore{\hskip -0.05 in
        \subfigure[]{
            \includegraphics[width=0.3\textwidth]{Combination_27_31_modes_old_new_calc.eps}
        }
\hskip -0.13 in
        \subfigure[]{
            \includegraphics[width=0.3\textwidth]{Combination_27_22_modes.eps}
        }
\hskip -0.13 in
        \subfigure[]{
            \includegraphics[width=0.3\textwidth]{Combination_22_31_modes.eps}
        }
}
            \end{center}
\vspace{-0.5cm}
    \caption{Phase diagram in the space spanned by (a) A$_g$-27 and A$_g$-31 phonon modes, (b) A$_g$-27 and A$_g$-22 phonon modes, and (c) A$_g$-22 and A$_g$-31 phonon modes. The intensity is the gap size at $\Gamma$ point; the $-$ sign is added for the WTI phase to distinguish WTI (dark colors) and STI (bright colors). The boundary (zero-value) line is the Dirac topology surface.
    }
        \label{fig:Combination_Ag_27_31}
\end{figure*}

%\subsection{Phonon-space Dirac topology surface}
\emph{Phonon-space Dirac topology surface---}It is noteworthy that the Dirac point thus formed is fine tuned, not symmetry protected, as a 3D system cannot have a symmetry protected Dirac point at the $\Gamma$ point~\cite{ClassDiracYang_Nagaosa_NatureComm2014}. Hence, the Dirac point exists for only one Q value for a specific phonon mode. However, one can resort to different combinations of the A$_g$ modes for infinite possibilities of establishing the DSM state.

We further point out that since $n$ phonon modes can form a $n$-dimensional space, there exists an $n-1$ dimensional Dirac topology surface in this space where the DSM states live, separating STI and WTI located on the opposite sides of the surface. In Fig.~\ref{fig:Combination_Ag_27_31}, we use the gap-size plot to illustrate the Dirac topology surface (here it is a line) in the 2D space formed by two A$_g$ phonon modes. In the space of  A$_g$-27 and A$_g$-31 phonon modes as shown in Fig.~\ref{fig:Combination_Ag_27_31}(a), we find that the system goes through the TPT for different linear combinations of these two modes. The line separating the WTI to STI phase can be approximated by the linear equation  $10y+2.5x+0.75=0$. On the other hand, Fig.~\ref{fig:Combination_Ag_27_31}(c) shows that the Dirac line in the space spanned by the A$_g$-31 and A$_g$-22 modes is approximately independent of the A$_g$-22 mode. This seems to be consistent with the result that A$_g$-22 mode alone does not cause a TPT. However, as shown in Fig.~\ref{fig:Combination_Ag_27_31}(b), the A$_g$-22 mode together with a weak A$_g$-27 mode can drive the system to the WTI phase. The concept of phonon-space Dirac topology surface can be generalized to the 6D space of the A$_g$ phonon modes. This yields infinitely many ways of driving TPT in this system and an encouraging prospective considering the 31D space of all the optically active phonon modes.

%\section{Effective Hamiltonian}
\emph{Effective Hamiltonian---}The four low-energy states at $\Gamma$ point are two Kramers pairs formed by the Te2 and Te3 $5p$ orbitals. The essential low-energy physics can be described by the $\mathbf{k} \cdot \mathbf{p}$ model of Chen \textit{et. al.} ~\cite{ChenEffectiveH_PRL_2015}, which is nothing but the relativistic Dirac Hamiltonian:
\begin{equation}
    H(\mathbf{k}) = m\tau^x + v_xk_x\tau^z\sigma^y + v_yk_y\tau^z\sigma^x + v_zk_z\tau^y,
    \label{eqn:H_kp}
\end{equation}
where $\mathbf{\sigma}$ is the Pauli matrices acting on the spin components of the Kramers pairs and $\mathbf{\tau}$ the Pauli matrices acting on the valley or ``orbital'' components of the Kramers pairs. Note that the Hamiltonian is expressed in the following coordinate system: $\mathbf{v}$, $\mathbf{k}$ and $\sigma$’s $x$, $y$, $z$-axes correspond to crystal $a$, $b$, $c$-axes respectively, but $\tau$’s $x$, $y$, $z$-axes are rotated to correspond to crystal $b$, $c$, $a$-axes in favor of deriving Eq.~(\ref{eqn:Dirac}).
The main effect of the A$_g$ phonon modes is changing the mass $m$. As shown in Fig.~\ref{fig:Bandgap_displacement_summary}, for five out of the six symmetry protecting $A_g$ phonon modes the mass term in Eq.~(\ref{eqn:H_kp}) depends linearly on the deformation $Q_i$ and for the $i$th mode there is a value of deformation $Q_i^0$ where it changes sign, indicating the STI-WTI phase transition. The contributions from these modes add mass linearly [Fig.~\ref{fig:Combination_Ag_27_31}(a)] such that
\begin{equation}
m \simeq \sum_i A_i (Q_i - Q_i^0).
\label{eqn:linear}
\end{equation}
By contrast, the mass change induced by the A$_g$-22 mode is of the quadratic form $m = m_0 + B Q^2$ and the effects of its combination with the other A$_g$ modes are more complicated.
%, as shown in  Fig.~\ref{fig:Combination_Ag_27_31}(b) and \ref{fig:Combination_Ag_27_31}(c).

An interesting situation emerges when the mass term $m$ changes sign at some interface, for instance on the plane   $z=z_0$ (crystal $c$-axis)~\cite{Volkov1985,Kusmartsev1985}. Then the eigenvalue equation for Hamiltonian (\ref{eqn:H_kp}) can be written as
\begin{eqnarray}
 && \left(
 \begin{array}{cc}
 -E +h_{\perp}(k_{\perp}) & m(z) + v_z\frac{d}{dz}\\
 m(z) - v_z\frac{d}{dz} & -E - h_{\perp}(k_{\perp})
 \end{array}
 \right) \left(
 \begin{array}{c}
 \psi_R\\
 \psi_L
 \end{array}
 \right) =0, \nonumber\\
 &&  h_{\perp}(k_{\perp}) = v_xk_x\sigma^y + v_yk_y\sigma^x.
 \label{eqn:Dirac}
 \end{eqnarray}
 Depending on whether $m(z)$ behaves as sign$(z-z_0)$ or as $-$sign$(z-z_0)$, this equation has a solution in the form of a single Weyl mode $E = \pm h_{\perp}(k_{\perp})$ bound to the surface with the wave function (for $m(+\infty) >0$)
 \begin{eqnarray}
 \Psi(z) = \left(
 \begin{array}{c}
 0\\
 1
 \end{array}
 \right) \exp\Big[- v_z^{-1}\int_{0}^{z-z_0} m(\xi) d\xi\Big].
 \end{eqnarray}
 Since the A$_g$ modes do not break the crystallographic symmetry, the metastable state created by laser pumping will likely consist of domains with different signed mass.
 Such domains will have the Weyl modes of opposite chirality propagating along the domain boundaries. This is a major effect of the A$_g$-mode distortions.
 %In this paper we consider lattice distortions originating from phonons; we will treat these distortions as adiabatic assuming that due to significant lattice anharmonism they are long lived. %One can imagine a situation when significant lattice anharmonicity produce long living domains with deformations of opposite sign.

%\section{Conclusion and Outlook}
%\label{Sec:Conclusion}
In summary, we have found that the atomic displacements corresponding to five out of six A$_g$ Raman-active phonon modes and their combinations can drive ZrTe$_5$ crystal from STI to WTI thereby forming a Dirac cone at the phase transition point. With more modes not breaking the crystallographic symmetry, the metastable state created by the laser pumping will likely consist of domains with different phases and we predict that such domains will have the Weyl modes of opposite chirality propagating along the domain boundaries. Studying phonon-space topology surfaces provides a new generic route to understanding and utilizing the exotic physical properties of ZrTe$_5$ and related quantum materials, e.g., finding transient transition to a Weyl semimetal state in {\mater} via photoexcited infrared-active phonon modes. We anticipate that our results will encourage more work in the field of ultrafast TPT.

%\section{Acknowledgements}
This work was supported by U.S. Department of Energy (DOE) the Office of Basic Energy Sciences, Materials Sciences and Engineering Division under Contract No. DE-SC0012704. X.J. acknowledges the visiting scholarship of Brookhaven National Laboratory and the financial support of China Scholarship Council.

%\bibliography{references}
%merlin.mbs apsrev4-1.bst 2010-07-25 4.21a (PWD, AO, DPC) hacked
%Control: key (0)
%Control: author (72) initials jnrlst
%Control: editor formatted (1) identically to author
%Control: production of article title (-1) disabled
%Control: page (0) single
%Control: year (1) truncated
%Control: production of eprint (0) enabled
%

\end{document}